\documentstyle[preprint,aps,epsf,eqsecnum,mbf]{revtex}
\tightenlines
\def\eq#1{Eq.\ (\ref{#1})}
\def\gtrsim{\begin{array}{c} > \\ [-3mm] \sim \\ \end{array}}

\def\bp{{\mbf p}}

\def\bfsigma{{\mbf \sigma}}

\def\CP{{\cal P}}

\def\SM3{\Sigma N (3/2)}
\def\SN1{\Sigma N (1/2)}

\def\TS1{\hbox{}^3S_1}
\def\TD1{\hbox{}^3D_1}

\begin{document}
\draft
\preprint{KUNS-1725/June 2001}

\title   {
Interactions between Octet Baryons in the $\mib{SU_6}$ Quark Model
         }

\author  {
Y. Fujiwara$^1$, M. Kohno$^2$, C. Nakamoto$^3$,
and Y. Suzuki$^4$
         }
\address {
$^1$Department of Physics, Kyoto University,
Kyoto 606-8502, Japan \\
$^2$Physics Division, Kyushu Dental College,
Kitakyushu 803-8580, Japan \\
$^3$Suzuka National College of Technology,
Suzuka 510-0294, Japan \\
$^4$Department of Physics,
Niigata University, Niigata 950-2181, Japan
         }
\maketitle

\bigskip
\bigskip

\begin{abstract}
Baryon-baryon interactions for the complete baryon octet ($B_8$)
are investigated in a unified framework of the resonating-group method,
in which the spin-flavor $SU_6$ quark-model wave functions are employed.
Model parameters are determined to reproduce properties
of the nucleon-nucleon system and the low-energy
cross section data for the hyperon-nucleon interaction.
We then proceed to explore $B_8 B_8$ interactions
in the strangeness $S=-2,~-3$ and $-4$ sectors.
The $S$-wave phase-shift behavior and total cross sections are
systematically understood by 1) the spin-flavor $SU_6$ symmetry,
2) the special role of the pion exchange
and 3) the flavor symmetry breaking.
\end{abstract}

\bigskip


\section{Introduction}

In the quark model, the baryon-baryon interactions for the complete
octet baryons ($B_8=N,~\Lambda,~\Sigma$ and $\Xi$) are treated
entirely equivalently with the well-known
nucleon-nucleon ($NN$) interaction.
Once the quark-model Hamiltonian is assumed in the framework
of the resonating-group method (RGM), the explicit evaluation of the
spin-flavor factors leads to the stringent flavor
dependence appearing in various interaction pieces.
We can thus minimize the ambiguity of the model parameters
by utilizing the rich knowledge of the $NN$ interaction.

In this study we first upgrade our previous
model \cite{FSS} for the $NN$ and hyperon-nucleon ($YN$) interactions
by incorporating more complete effective meson-exchange potentials (EMEP)
such as the vector mesons and some extra interaction pieces.
This model is named fss2 \cite{fss2} after the pioneering model FSS.
Fixing the model parameters in the strangeness $S=0$ and $-1$
sectors, we proceed to explore the $B_8 B_8$ interactions
in $S=-2,~-3$ and $-4$ sectors. These include the $\Lambda \Lambda$ and
$\Xi N$ interactions, which have recently been attracting
much interest in the rapidly developing fields of hypernuclear physics
and strangeness nuclear matter. 

In the next section, we recapitulate the formulation
of the $(3q)$-$(3q)$ Lippmann-Schwinger RGM \cite{LSRGM}.
In Sec.\,III, we summarize the essential features
of the $NN$ and $YN$ interactions, in order to furnish the basic
components to understand the phase-shift behavior
of the $B_8 B_8$ interactions in a unified way.
The model predictions of the $B_8 B_8$ interactions
in the $S=-2,~-3$ and $-4$ sectors are given in Sec.\,IV,
with respect to the $S$-wave phase shifts and the total cross
sections. The final section is devoted to a summary.

\section{Formulation}

The quark-model Hamiltonian $H$ consists of the phenomenological
confinement potential $U^{\rm Cf}_{ij}$, the colored
version of the full Fermi-Breit (FB) interaction $U^{\rm FB}_{ij}$
with explicit quark-mass dependence,
and the EMEP $U^{\Omega \beta}_{ij}$ generated from 
the scalar ($\Omega$=S), pseudoscalar (PS) and vector (V)
meson exchange potentials acting between quarks:
\begin{eqnarray}
H=\sum^6_{i=1} \left( m_ic^2+{\bp^2_i \over 2m_i}-T_G \right)
+\sum^6_{i<j} \left( U^{\rm Cf}_{ij}+U^{\rm FB}_{ij}
+\sum_\beta U^{{\rm S}\beta}_{ij}
+\sum_\beta U^{{\rm PS}\beta}_{ij}
+\sum_\beta U^{{\ V}\beta}_{ij} \right)\ .
\label{fm1}
\end{eqnarray}
It is important to include the momentum-dependent Bryan-Scott
term \cite{BR67} in the S- and V-meson contributions, 
in order to remedy the shortcoming of our previous model FSS,
namely the single-particle (s.p.) potential in nuclear matter
is too attractive in the high-momentum region $k \gtrsim 6~\hbox{fm}^{-1}$.
Another important feature of the present model is the introduction
of vector mesons for improving the fit to the $NN$ phase-shift parameters.
Since the dominant effect of the $\omega$-meson
repulsion and the $LS$ components of $\rho$, $\omega$ and $K^*$ mesons
are already accounted for by the FB interaction,
only the quadratic $LS$ component of the octet mesons
is expected to play an important role in partially canceling
the strong one-pion tensor force.
Further details of the model fss2 are given in \cite{fss2}.
The model parameters are fixed to reproduce the most recent results
of the phase shift analysis SP99 \cite{SAID} for the $np$ scattering
with partial waves $J \leq 2$ and incident
energies $T_{\rm lab} \leq 350~\hbox{MeV}$,
under the constraint of the deuteron binding energy
and the $\hbox{}^1S_0$ $NN$ scattering length, as well as the
low-energy $YN$ total cross section data.
Owing to the introduction of the vector mesons,
the model fss2 in the $NN$ sector has attained an accuracy
almost comparable to that of one-boson exchange potential (OBEP) models.  
For example, the $\chi^2$ values defined by $\chi^2=\sum_{i=1}^N
\left( \delta_i^{cal}-\delta_i^{exp}\right)^2/N$ for
the $J \leq 2$ phase-shift parameters in the energy
range $T_{\rm lab}=25$ - 300 MeV are $\sqrt{\chi^2}$
= $0.59^\circ$, $1.10^\circ$, $1.40^\circ$ and $1.32^\circ$ for
fss2, OBEP, Paris and Bonn, respectively.
The existing data for the $YN$ scattering are well reproduced
and the essential feature of the $\Lambda N$-$\Sigma N$ coupling
is almost unchanged from our previous models.

The two-baryon systems composed of the complete baryon octet are
classified as
\begin{eqnarray}
1/2(11) \times 1/2(11)=\{0,~1\}\,\{(22)+(30)+(03)+(11)_s
+(11)_a+(00)\} \ ,
\label{fm3}
\end{eqnarray}
where $S(\lambda \mu)$ stands for the spin value $S$ and
the flavor $SU_3$ representation label $(\lambda \mu)$.
If the space-spin states are classified by the flavor exchange
symmetry $\CP$, the correspondence between the $SU_3$ basis
and the isospin basis becomes very transparent,
as in Table \ref{table1}.
This correspondence is essential in the following discussion.
The key point is that the quark-model
Hamiltonian \eq{fm1} is approximately $SU_3$ scalar.
If we ignore the EMEP terms $U^{{\rm S,PS,V}\beta}_{ij}$,
this is fairly apparent since the flavor dependence
appears only through the moderate mass difference of the
up-down and strange quarks. In the FB interaction $U^{\rm FB}_{ij}$,
this is a direct consequence of the fact that the
gluons do not have the flavor degree of freedom.
In the EMEP terms, the $SU_3$ scalar property of the interaction
is not apparent since the mesons have
the flavor degree of freedom. Nevertheless, one can easily show
that the interaction Hamiltonian is actually $SU_3$ scalar   
if the masses of the octet mesons are all equal within each of the S,
PS and V mesons.
A nice feature of the quark model is that the approximate $SU_3$-scalar
property of the total Hamiltonian is automatically incorporated
in the model. On the other hand, in the OBEP a similar situation is only
realized by assuming the $SU_3$ relations for the many baryon-meson
coupling constants.
If the Hamiltonian is exactly $SU_3$ scalar, the $SU_3$ states
with a common $(\lambda \mu)$ in Table \ref{table1} should have the same
baryon-baryon interaction.
For example, the same (22) symmetry appears
in several $\hbox{}^1S_0$ states; i.e., $NN (I=1)$, $\Sigma N (I=3/2)$,
$\Sigma \Sigma (I=2)$, $\Xi \Sigma (I=3/2)$ and $\Xi \Xi (I=1)$.
The $\hbox{}^1S_0$ phase shifts of these channels have
very similar behavior, as is shown in Fig.\,\ref{fig1}.
In reality, the $SU_3$ symmetry is broken, but in a very specific way.
The mechanism of the flavor symmetry breaking (FSB) depends on the
details of the model. In the present framework, the following
three factors cause FSB:
\begin{enumerate}
\setlength{\itemsep}{-2mm}
\item[1)] The strange to up-down quark mass ratio $\lambda= 
m_s/m_{ud}=1.551~(\hbox{for~fss2}) > 1$ in the kinetic-energy
term and $U^{\rm FB}_{ij}$.
\item[2)] The singlet-octet meson mixing in $U^{{\rm S,PS,V}\beta}_{ij}$.
\item[3)] The meson and baryon mass splitting
in $U^{{\rm S,PS,V}\beta}_{ij}$ and the kinetic-energy term,
and the resultant difference of the threshold energies.
\end{enumerate}

\section{Characteristics of the $\mib{NN}$ and $\mib{YN}$ Interactions
and the Basic Viewpoint}

\begin{figure}[h]
\begin{minipage}[h]{0.46\textwidth}
\epsfxsize=\textwidth
\epsffile{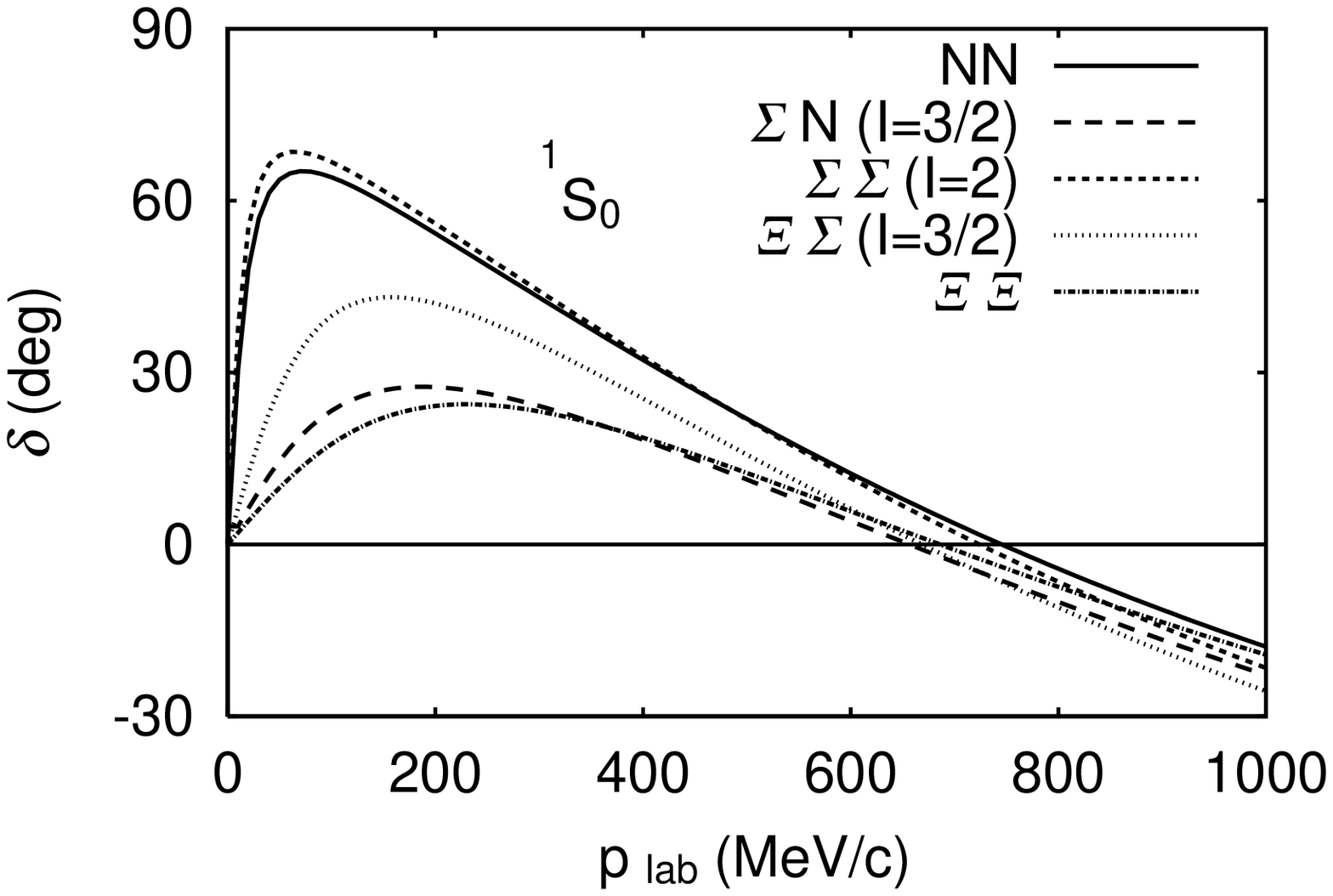}
\caption{
$\hbox{}^1S_0$ phase shifts for the $B_8$-$B_8$ interactions
with the pure (22) state.
}
\label{fig1}
\end{minipage}
\hfill
\begin{minipage}[h]{0.46\textwidth}
\epsfxsize=\textwidth
\epsffile{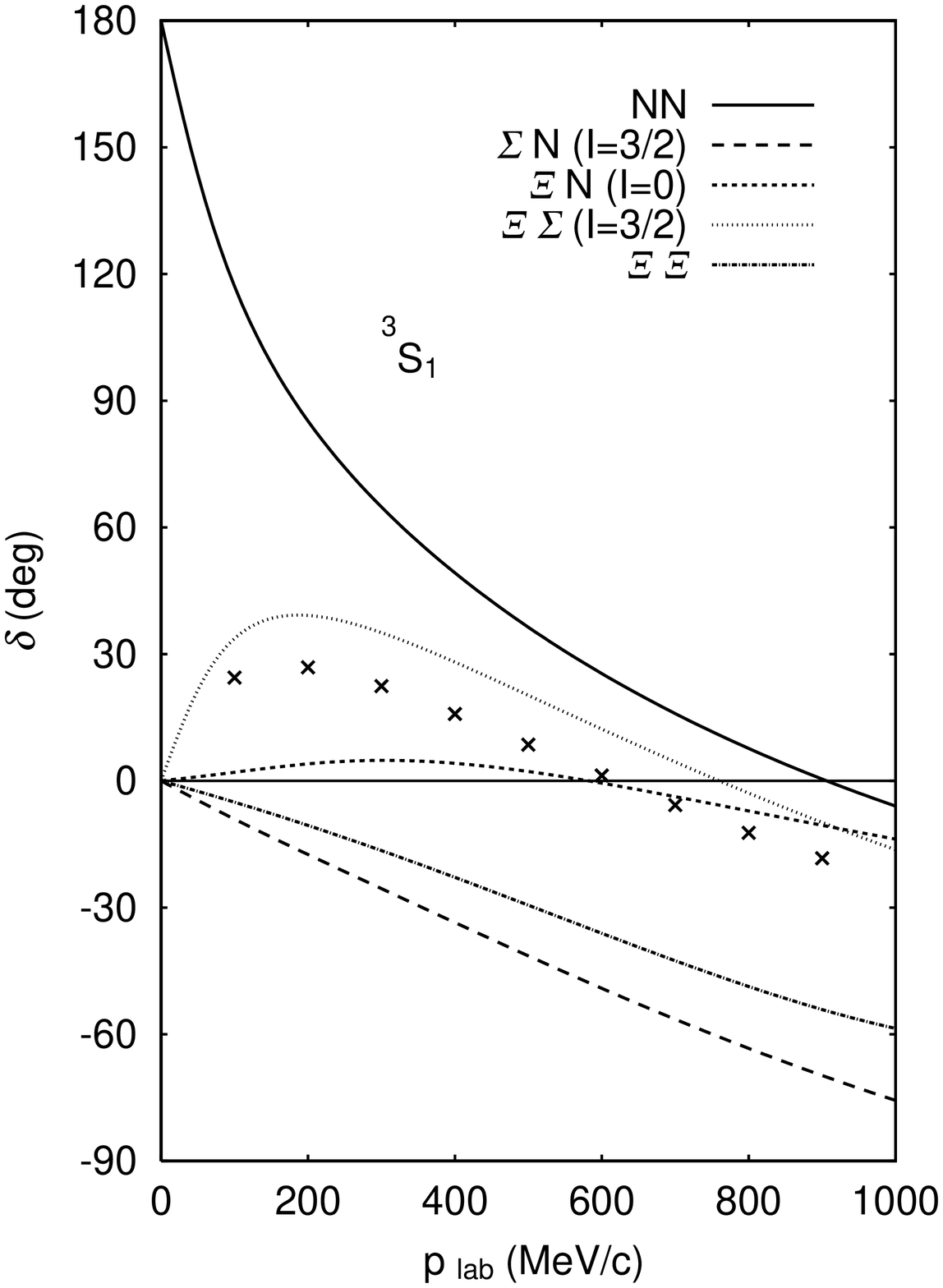}
\caption{$\hbox{}^3S_1$ phase shifts for the (03) ($NN$,
$\Xi \Sigma (I=3/2)$), $(11)_a$ ($\Xi N (I=0)$),
and (30) ($\Sigma N (I=3/2)$, $\Xi \Xi$) states.
The $\hbox{}^3S$ phase shift predicted only by the $NN$ central
interaction is also shown by crosses.
} 
\label{fig2}
\end{minipage}
\end{figure}

The approximate $SU_3$-scalar property of the Hamiltonian
implies that accurate knowledge of the $NN$ and $YN$ interactions
is crucial to understand the $B_8$-$B_8$ interactions
in the $S=-2,~-3$ and $-4$ sectors.
The following four points concerning the qualitative behavior
of the $NN$ and $YN$ interactions are essential. \cite{FSS,fss2}
First, in the $NN$ system the $\hbox{}^1S_0$ state
with isospin $I=1$ consists of the pure (22) state,
and the phase shift shows a clear resonance behavior
reaching more than $60 ^\circ$ (see Fig.\,\ref{fig1}).
On the other hand, the $\hbox{}^3S_1$ state with $I=1$ is composed
of the pure (03) state, and the deuteron is bound in this channel
owing to the strong one-pion tensor force.
If we switch off this strong tensor force, the $\hbox{}^3S$ phase shift
rises to $20^\circ \sim 30^\circ$ at most,
indicating that the central attraction of the (03) state is not as strong
as that of the (22) state (see crosses in Fig.\,\ref{fig2}).
Detailed analysis of the $YN$ interaction has
clarified that the $(11)_s$ state for the $\hbox{}^1S_0$ state
and the (30) state for the $\hbox{}^3S_1$ state are both 
strongly repulsive, reflecting that the most compact $(0s)^6$ configuration
is completely Pauli forbidden for the $(11)_s$ state
and almost forbidden for the (30) state. \cite{NA95}
The repulsive behavior of the $\Sigma N (I=3/2)$ $\hbox{}^3S_1$ state
with the pure (30) symmetry should be observed as a strong isospin
dependence of the $\Sigma$ s.p. potential.
On the other hand, the experimental evidence of the repulsion
is not clear for the $\Sigma N (I=1/2)$ $\hbox{}^1S_0$ state,
which contains 90 \% $(11)_s$ component.
This is because the observables are usually composed of the contributions
both from the $\hbox{}^1S_0$ and $\hbox{}^3S_1$ states,
and the $\hbox{}^3S_1$ state in this channel involves a rather
cumbersome $\Lambda N$-$\Sigma N (I=1/2)$ channel coupling.
Nevertheless, the repulsive character of the $(11)_s$ state
is not inconsistent with the present experimental evidence,
in the sense that FSS and fss2 reproduce the available low-energy
cross section data of the $\Lambda p$ and $\Sigma^- p$ scatterings
quite well.
The strength of this repulsion depends on the detailed framework
of the quark model.
In the OBEP approach, even the qualitative features of these
interactions are sometimes not reproduced.
For example, in the $\Sigma N (I=3/2)$ $\hbox{}^3S_1$ state almost all
the Nijmegen soft-core models \cite{NSC89,NSC97} predict
a broad resonance around the intermediate energy region
of $p_\Sigma \sim 400\,\hbox{-}\,600$ MeV,    
although the low-energy behavior of the phase shifts are surely
repulsive. 

The following two additional features of the present model should be
kept in mind during the discussion of the $B_8 B_8$ interactions
in the $S=-2,~-3$ and $-4$ sectors.
First, the flavor-singlet (00) state which appears in the $S=-2$ sector
for the first time is usually attractive in the quark model,
owing to the $(\bfsigma_1 \cdot \bfsigma_2)
(\lambda^C_1 \lambda^C_2)$-type color-magnetic interaction
involved in $U^{\rm FB}_{ij}$.
Whether the $H$-dibaryon state with the pure (00) is bound or not
depends on how much the strangeness-exchange EMEP contribution 
cancels the strong channel coupling effect from
the FB interaction $U^{\rm FB}_{ij}$. \cite{NA97}
Next, the $(11)_a$ configuration which appears in the $S=-1$ sector
partially appears in the $\hbox{}^3S_1$ state in $\Xi N (I=0)$ in
the pure $SU_3$ form.
The fss2 prediction for the phase shift of this interaction,
depicted in Fig.\,\ref{fig2}, implies that this interaction 
is very weak, since the phase-shift rise is only $5^\circ$.
Nevertheless, the phase shift behavior of the $\hbox{}^3S_1$ states
for the $\Lambda N$ and $\Sigma N (I=1/2)$ channels
in the $S=-1$ sector is very different \cite{FSS,fss2}.
This is apparently due to the strong effect of the one-pion tensor force,
which is present in the $\Sigma N$ channel,
while absent in the $\Lambda N$ channel. 
We can therefore conclude that a further important factor besides the FSB
is the specific effect of the Goldstone-boson pions,
which is very much channel dependent.
As a consequence of the $SU_3$ relations,
the role of the pion is generally reduced if the strangeness
involved in the system increases.

\section{Results in $\mib{S}=-2,~-3$ and $-4$ Sectors}

\begin{figure}[h]
\begin{minipage}[h]{0.46\textwidth}
\epsfxsize=\textwidth
\epsffile{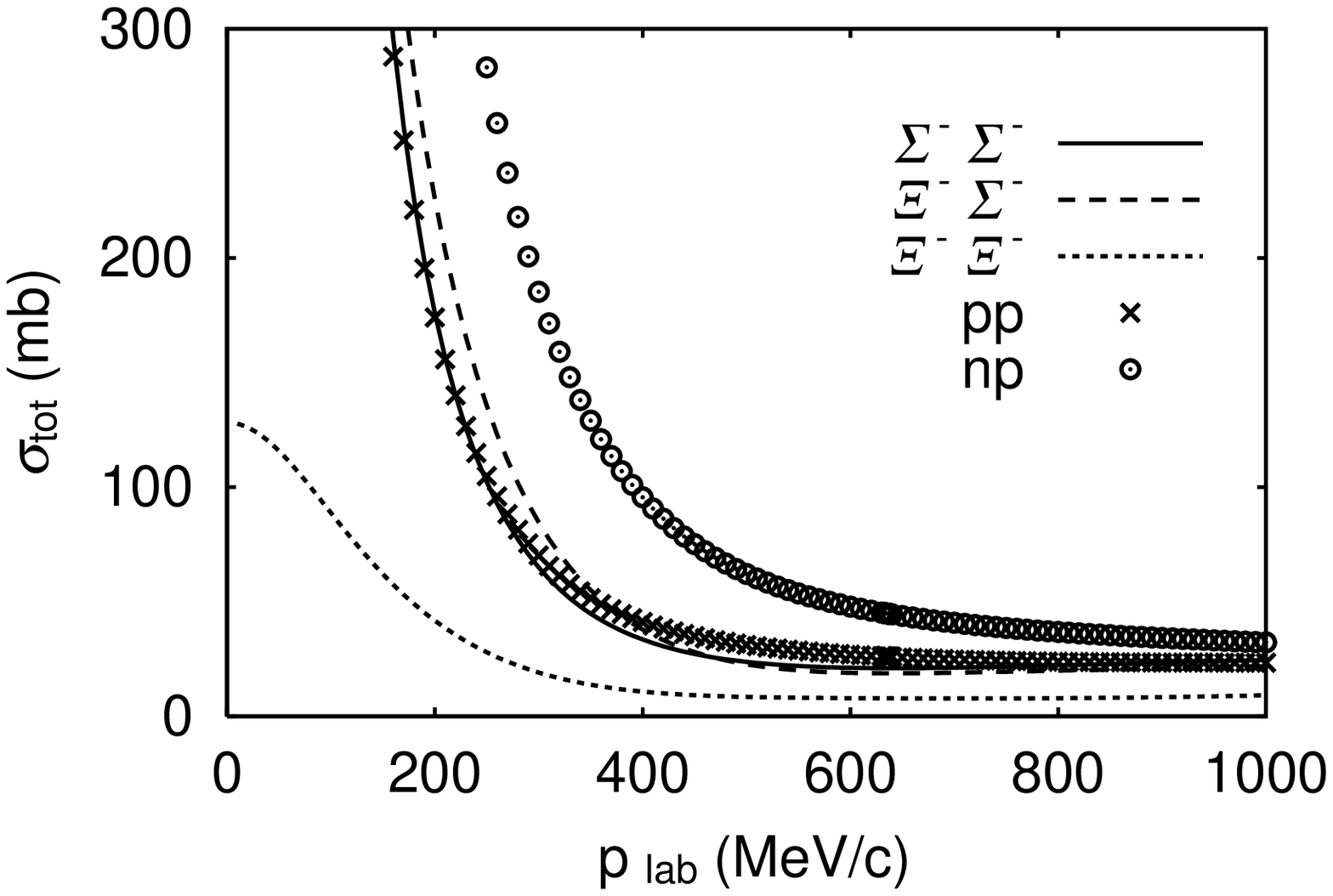}
\caption{
Total elastic cross sections for the
pure (22) state ($pp$, $\Sigma^- \Sigma^-$, $\Xi^- \Xi^-$) and
for the (22)+(03) states ($np$, $\Xi^- \Sigma^-$).
}
\label{fig3}
\end{minipage}
\hfill
\begin{minipage}[h]{0.46\textwidth}
\epsfxsize=\textwidth
\epsffile{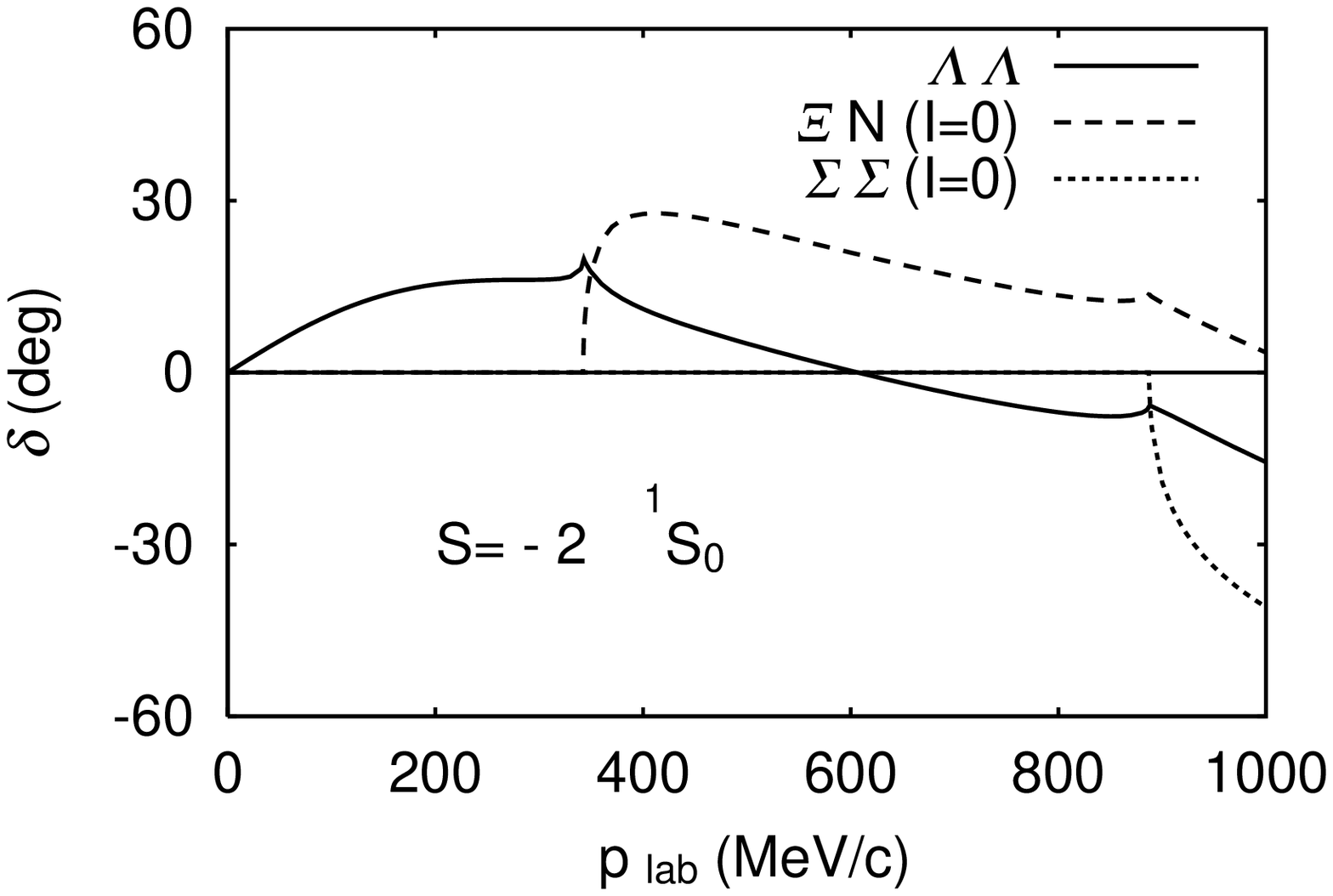}
\caption{$\hbox{}^1S_0$ phase shifts
in the $\Lambda \Lambda$-$\Xi N$- $\Sigma \Sigma$ coupled-channel
system with $I=0$.
}
\label{fig4}
\end{minipage}
\end{figure}

Table \ref{table1} shows a ``reflection'' symmetry
with respect to the interchange between $S=0$ and $S=-4$ sectors,
and also between $S=-1$ and $S=-3$ sectors.
Just like the particle-hole symmetry in the nuclear shell model,
the $SU_3$ state on one side is obtained from the $(\lambda \mu)$ state
on the other side by simply interchanging the $\lambda$ and $\mu$.
Since the (22) (and also $(11)_s$) symmetry in the $\hbox{}^1S_0$ state
returns to itself, the $NN$, $\Lambda N$ and $\Sigma N$ interactions
with $N$ being replaced by $\Xi$ should be very similar to the
original ones.
On the other hand, in the $\hbox{}^3S_1$ state the (03) symmetry changes
into the (30) symmetry and the attractive interaction turns to the
repulsive one. For example, $\Xi \Xi (I=0)$ interaction
with the pure (30) symmetry is repulsive.
Figure \ref{fig3} shows that the $\Xi \Xi$ total cross sections
are about 1/4 $\sim$ 1/5 of the $NN$ cross sections.
This result is different from the very large prediction of the Nijmegen
soft-core potentials in \cite{ST99},
which have strong attractions in all the (22) channels
except for the $NN$ and $\Sigma^+ p$ ones.
Among the strangeness $B_8$-$B_8$ channels
having the $\hbox{}^1S_0$ (22) configuration,
the most attractive $\hbox{}^3S_1$ state is expected
for the $\Xi \Sigma (I=3/2)$ interaction.
Table \ref{table1} shows that this interaction has a very interesting
feature: the two $SU_3$ symmetries (22) and (03) in
the $NN$ interaction appear in a common isospin state $I=3/2$.
Unlike the (03) state in $NN$, the (03) state in $\Xi \Sigma (I=3/2)$ is
only moderately attractive, since the $\Xi \Sigma$ system does not allow
the strong one-pion exchange in the exchange Feynman diagram.
Even in the direct Feynman diagram, the inclusion of the strangeness
reduces the one-pion exchange effect drastically
through the $SU_3$ relations.
The $\Xi^- \Sigma^-$ interaction thus gives the largest total cross
sections in the strangeness sector,
together with the $\Sigma^- \Sigma^-$ interaction,
as seen in Fig.\,\ref{fig3}. The magnitudes of these total cross sections,
however, are at most comparable with the $pp$ total cross sections.
The $\Xi \Lambda$-$\Xi \Sigma (I=1/2)$ coupled-channel problem,
on the other hand, is less interesting, since the one-pion tensor force
in the $\hbox{}^3S_1+\hbox{}^3D_1$ state becomes less effective
due to the strong repulsion of the (30) component.
This is in contrast with the strong $\Lambda N$-$\Sigma N (I=1/2)$ channel
coupling, which leads to the well-known cusp structure
in the $\Lambda N$ total cross sections.

The baryon-baryon interactions in the $S=-2$ sector constitute
the most difficult case to analyze, involving three different
types of two-baryon configurations:
$\Lambda \Lambda$-$\Xi N$-$\Sigma \Sigma$ for $I=0$ and
$\Xi N$-$\Sigma \Lambda$-$\Sigma \Sigma$ for $I=1$.
In this case, the isospin dependence of the interaction is
very important, just as in the $\Sigma N (I)$ interactions
with $I=1/2$ and 3/2.
Figure \ref{fig4} shows the $\hbox{}^1S_0$ phase-shift behavior of
the full $\Lambda \Lambda$-$\Xi N$-$\Sigma \Sigma$ coupled-channel
system with $I=0$, in which the $H$-dibaryon bound state might exist.
In the previous model FSS the $\Lambda \Lambda$ phase shift rises
to $40^\circ$ \cite{NA00}, while in the present fss2 it rises only
to $\sim 20^\circ$ at most. The situation is the same as
in the $\Xi N (I=0)$ phase shift. It rises only
to $30^\circ \sim 40^\circ$ in fss2. Table \ref{table1} shows
that the largest contribution of the (00) component is realized not
in the $\Lambda \Lambda$ channel, but in the $\Xi N (I=0)$ channel.
This implies that the attractive effect of the (00) configuration
is smaller in fss2 than in FSS. Since FSS does not have the $H$-dibaryon
bound state \cite{NA00}, fss2 does not have it either.
As to the $\Lambda \Lambda$ interaction, it has been claimed that
a phase-shift rise on the order of $40^\circ$ is at least
necessary to explain the known three events
of the double $\Lambda$-hypernuclei.
However, the recently discovered the ``Demachi-Yanagi event'' \cite{IC00} for
$\hbox{}^{10}_{\Lambda \Lambda}\hbox{Be}$ and the ``Nagara event'' \cite{TA01} 
for $\hbox{}^{\ 6}_{\Lambda \Lambda}\hbox{He}$ indicate
that the $\Lambda \Lambda$ interaction is less attractive.
A rough estimate of $\Delta B_{\Lambda \Lambda}$
for $\hbox{}^{\ 6}_{\Lambda \Lambda}\hbox{He}$ in terms of the $G$-matrix
calculation using fss2 is about 1 MeV, which is consistent
with this experimental observation.

\begin{figure}[h]
\begin{minipage}{0.46\textwidth}
\epsfxsize=\textwidth
\epsffile{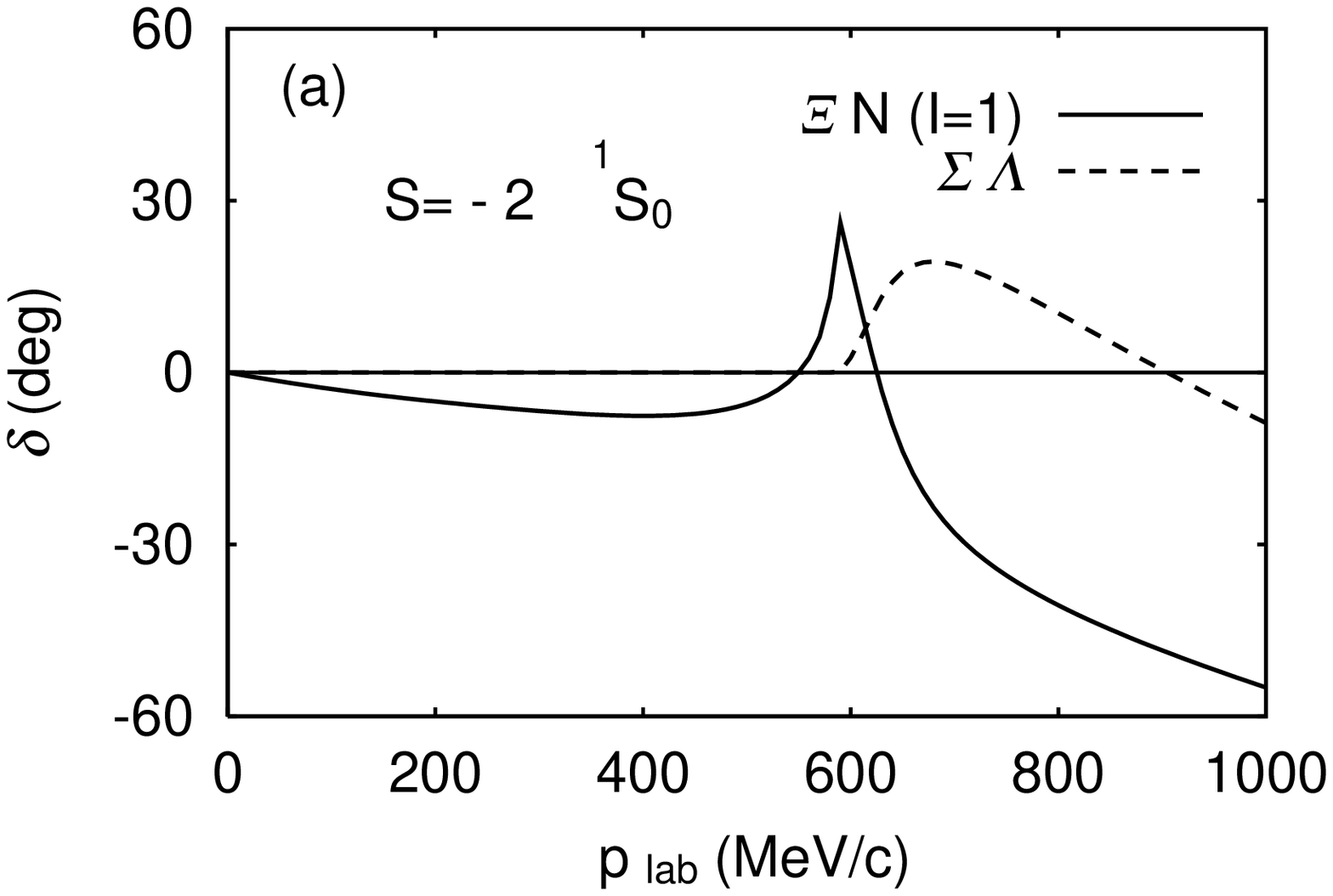}
\end{minipage}~%
\begin{minipage}{0.46\textwidth}
\epsfxsize=\textwidth
\epsffile{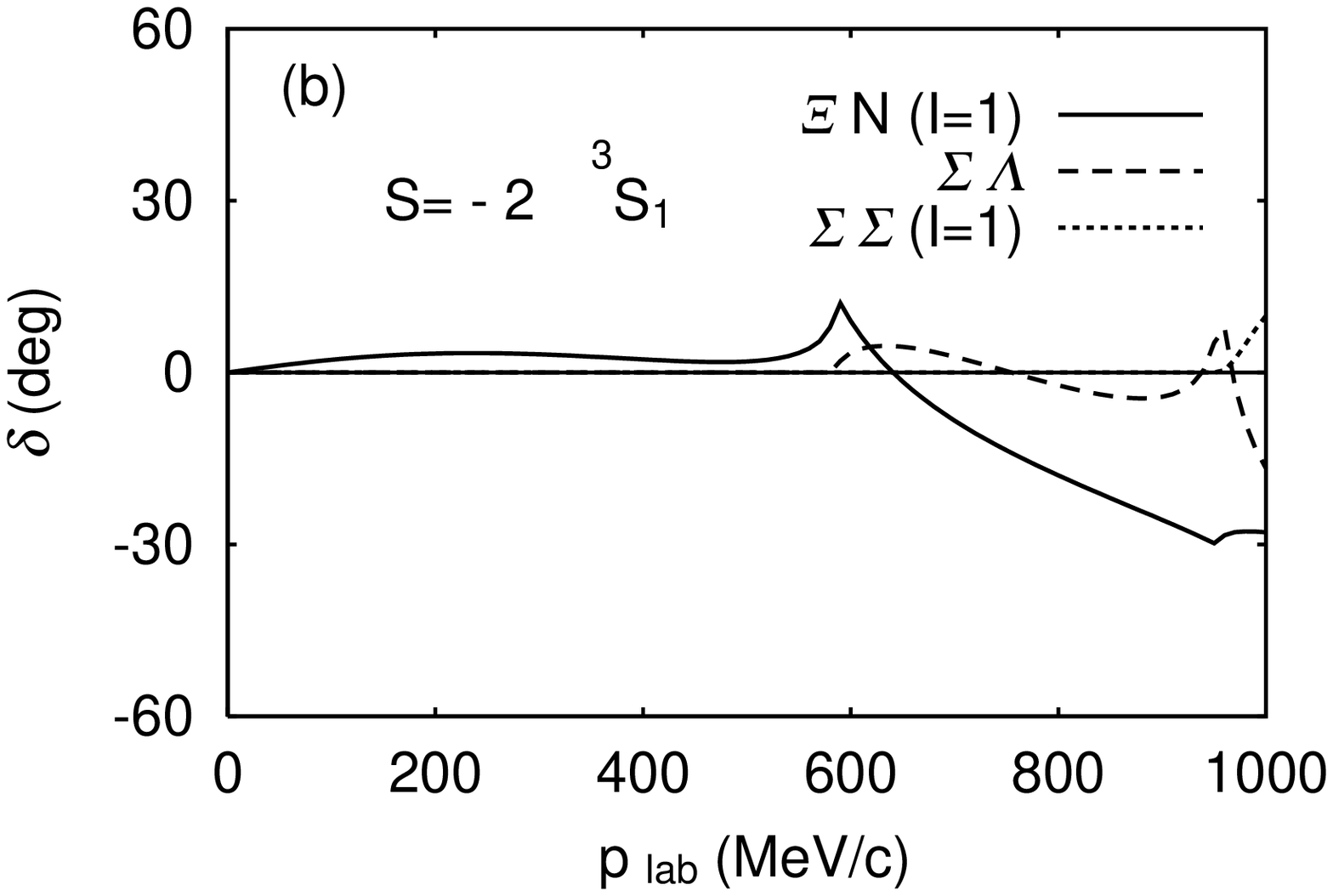}
\end{minipage}
\bigskip
\caption{
(a) $\hbox{}^1S_0$ phase shifts
in the $\Xi N$-$\Sigma \Lambda$-$\Sigma \Sigma$ coupled-channel
system with $I=1$. (b) The same as (a) but for the $\hbox{}^3S_1$ state.
}
\label{fig5}
\end{figure}

\begin{figure}[h]
\begin{minipage}{0.46\textwidth}
\epsfxsize=\textwidth
\epsffile{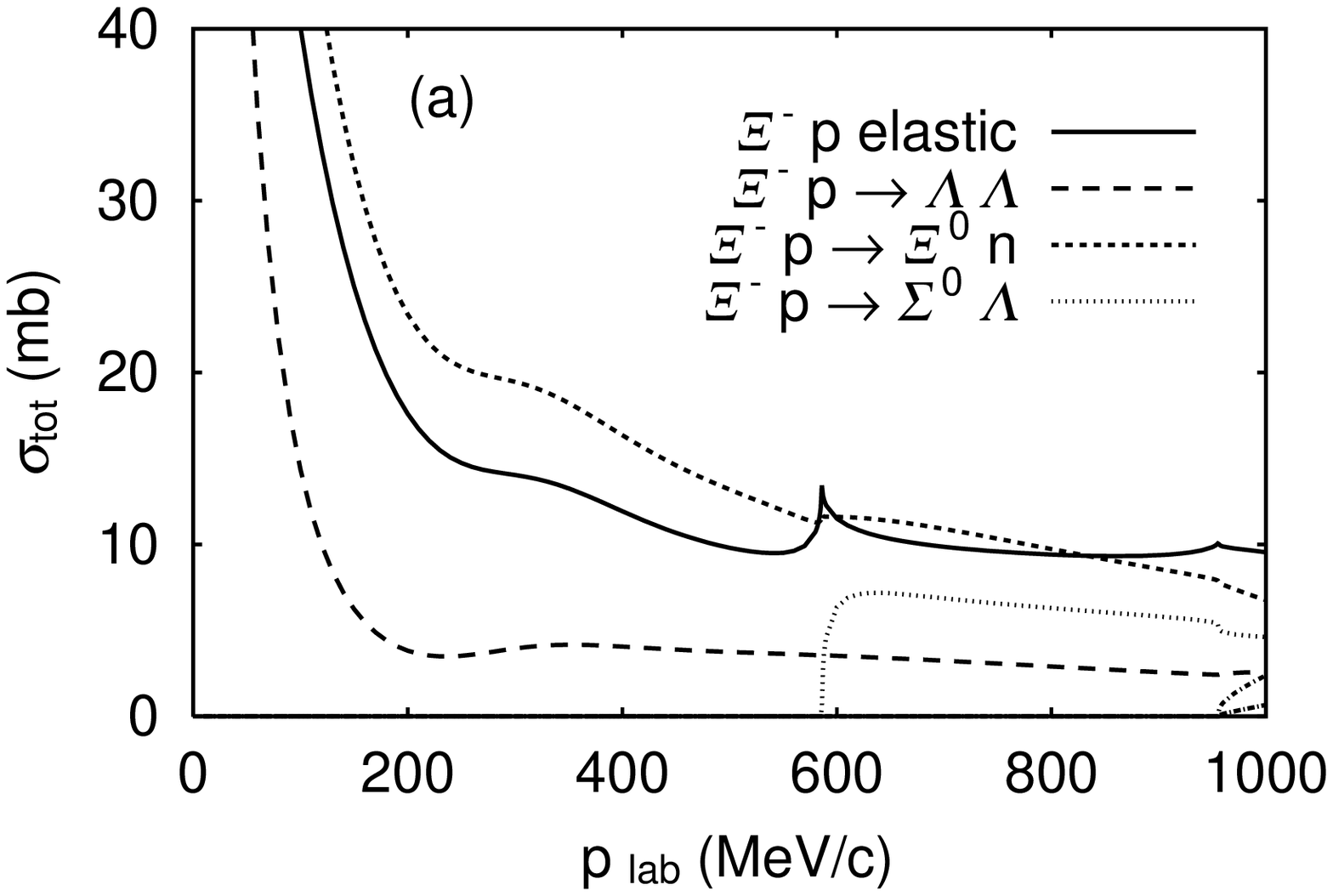}
\end{minipage}~%
\begin{minipage}{0.46\textwidth}
\epsfxsize=\textwidth
\epsffile{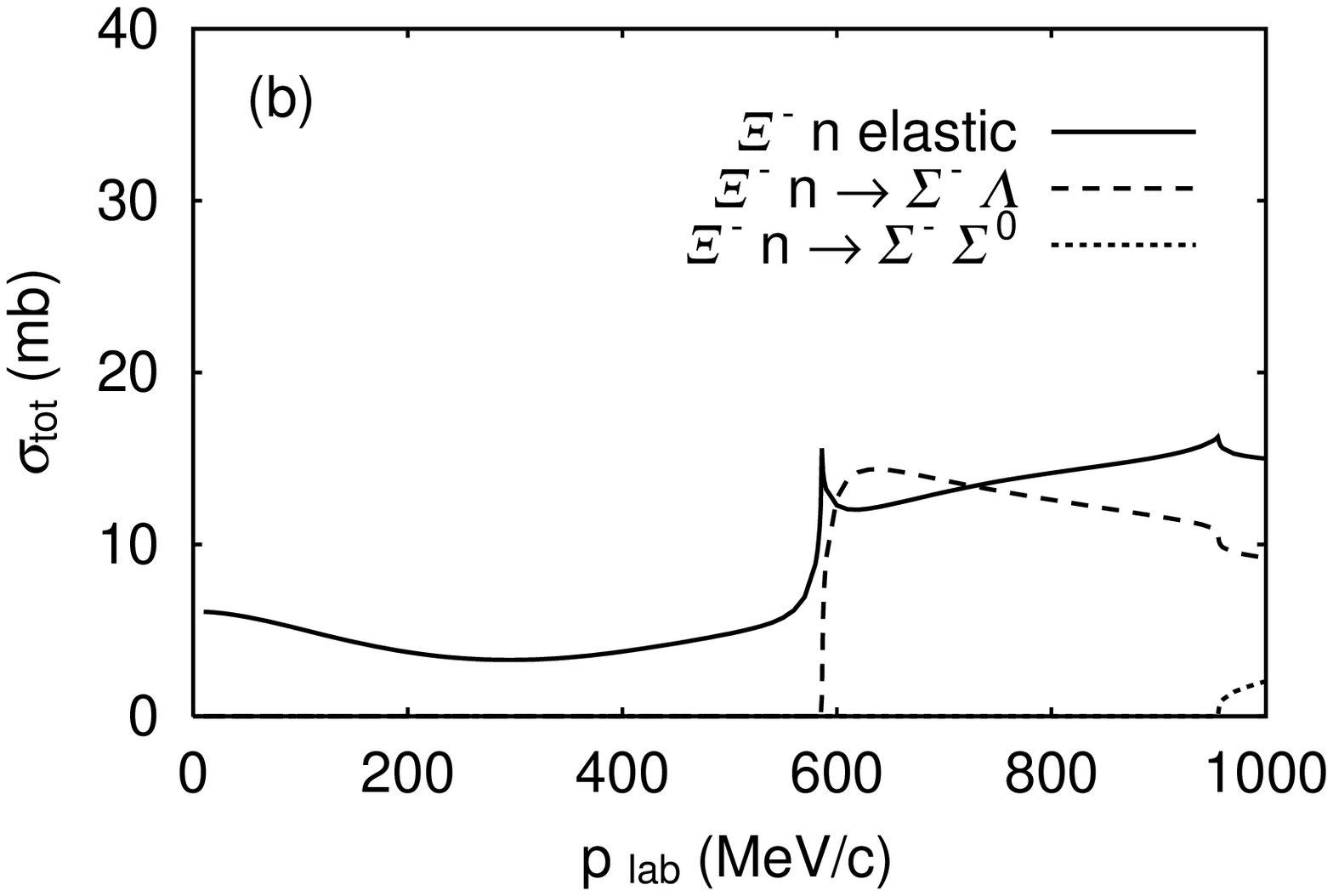}
\end{minipage}
\bigskip
\caption{(a) Total cross sections for $\Xi^- p$ scattering
with $I=0$ and 1 contributions. (b) The same as (a) but
for $\Xi^- n$ scattering only with $I=1$ contribution.
}
\label{fig6}
\end{figure}

In the isospin $I=1$ channel, the lowest incident baryon channel
in the $S=-2$ sector is the $\Xi N$ channel.
Figure \ref{fig5} shows the phase-shift behavior
of the $\hbox{}^1S_0$ and $\hbox{}^3S_1$ states, calculated for the
full coupled-channel system $\Xi N$-$\Sigma \Lambda$-$\Sigma \Sigma$ with
$I=1$.
In the $\Xi N (I=1)$ single-channel calculation, both of these
phase shifts show monotonic repulsive behavior, originating from
the main contributions of the $(11)_s$ and (30) components,
respectively. \cite{NA97}
In the full coupled-channel calculation, however, the channel coupling
effect between $\Xi N (I=1)$ and $\Sigma \Lambda$ channels is enhanced
by the cooperative role of the FB contribution $U^{\rm FB}_{ij}$ and
EMEP contribution $U^{\Omega \beta}_{ij}$ in the strangeness
exchange process.
As a result, the $\Xi N (I=1)$ phase shifts show very prominent 
cusp structure at the $\Sigma \Lambda$ threshold,
as seen in Fig.\,\ref{fig5}. Below the $\Sigma \Lambda$ threshold,
the phase-shift values are almost zero.
Subsequently, the $\Xi^0 p$ (and $\Xi^-n$) total cross sections
with the pure $I=1$ component are predicted to be very small 
below the $\Sigma \Lambda$ threshold
around $p_\Xi \sim 600~\hbox{MeV}/c$.
This behavior of the $\Xi^- n$ total cross sections,
illustrated in Fig.\,\ref{fig6}(b), is essentially the same
as the Nijmegen result in \cite{ST99}.
On the other hand, the $\Xi^- p$ total cross sections,
shown in Fig.\,\ref{fig6}(a), exhibit a typical channel-coupling
behavior similar to that of the $\Sigma^- p$ total cross sections.
These features demonstrate that the $\Sigma \Lambda$ channel-coupling
effect is very important for the correct description
of scattering observables, resulting in the strong
isospin dependence of the $\Xi N$ interaction. 

\section{Summary}

The conversion processes among octet baryons $B_8$ are most
straightforwardly incorporated in the coupled-channel formalism.
Since our quark model Hamiltonian is approximately $SU_3$ scalar,
it is crucial to clarify the characteristics
of the $B_8$-$B_8$ interaction for each of the $SU_3$ states,
rather than for each of the two-baryon systems.
An inter-baryon potential in a single baryon channel is sometimes used
for the study of hypernuclei and strangeness nuclear matter.
However, such effective interactions are very much
model dependent and linkage to the bare interactions like the ones
discussed here is sometimes obscured by inherent ambiguities
originating from the many-body calculations.
The widespread argument that the hypernuclear structure
reflects the hyperon-nucleon ($YN$) interaction rather faithfully
since the $YN$ interaction is weak should not be overemphasized. 

In this study we have upgraded our previous quark model \cite{FSS} for
the nucleon-nucleon ($NN$) and $YN$ interactions
by incorporating more complete effective meson-exchange potentials
such as the vector mesons and some extra interaction pieces.
This model fss2 \cite{fss2} reproduces the existing
data of the $NN$ and $YN$ interactions quite well.
We have then proceeded to predict all the $B_8 B_8$ interactions
in the strangeness $S=-2,~-3$ and $-4$ sectors, without adding
any extra parameters.
We have discussed some characteristic features of
the $B_8 B_8$ interactions, focusing on the qualitative aspect.
Among these features are the following: 1) There is no bound state
in the $B_8 B_8$ systems, except for the deuteron;
2) The $\Xi \Xi$ total cross sections are far smaller
than the $NN$ cross sections;
3) The $\Xi N$ interaction has a strong isospin dependence
similar to the $\Sigma N$ system;
4) The $\Xi^- \Sigma^-$ ($\Xi \Sigma (I=3/2)$) interaction is
moderately attractive.
The $S$-wave phase-shift behavior yielding these qualitative features
of the $B_8 B_8$ interactions is systematically understood
by 1) the spin-flavor $SU_6$ symmetry, 2) the special role
of the pion exchange and 3) the flavor symmetry breaking.
More detailed analysis of the $B_8 B_8$ interactions
predicted by the model fss2 and consistency with the available
experimental data will be published in a forthcoming paper.

\acknowledgments

This research is supported by Japan Grant-in-Aid for Scientific
Research from the Ministry of Education, Science, Sports and
Culture (12640265).

\begin{table}[h]
\caption{
The relationship between the isospin basis
and the flavor-$SU_3$ basis for the $B_8 B_8$ systems. 
The flavor-$SU_3$ symmetry is given by the Elliott
notation $(\lambda \mu)$. 
$\CP$ denotes the flavor exchange symmetry, and $I$ the isospin.}
\label{table1}
\vspace{0mm}
\begin{center}
\setlength{\tabcolsep}{6mm}
\begin{tabular}{c|c|c|c}
$S$ & $B_8\,B_8~(I)$ & ${\cal P}=+1$ (symmetric)
    & ${\cal P}=-1$ (antisymmetric) \\
\cline{3-4}
    & & $\hbox{}^1 E$ \quad or \quad $\hbox{}^3 O$
    & $\hbox{}^3 E$ \quad or \quad $\hbox{}^1 O$ \\ 
\hline
$0$ & $NN~(I=0)$ & $-$ & $(03)$  \\
  & $NN~(I=1)$ & $(22)$ & $-$ \\
\hline
  & $\Lambda N$ &  ${1 \over \sqrt{10}} [ (11)_s + 3 (22) ]$ &
${1 \over \sqrt{2}} [ -(11)_a + (03) ]$ \\
$-1$ & $\Sigma N~(I=1/2)$ & ${1 \over \sqrt{10}} [ 3 (11)_s - (22) ]$ &
${1 \over \sqrt{2}} [ (11)_a + (03) ]$ \\
  & $\Sigma N~(I=3/2)$ & $(22)$ & $(30)$ \\[1mm]
\hline
  &  $\Lambda\Lambda$ & $\frac{1}{\sqrt{5}}(11)_s
+\frac{9}{2\sqrt{30}}(22)+\frac{1}{2\sqrt{2}}(00)$ & $-$ \\
  & $\Xi N~(I=0)$ & $\frac{1}{\sqrt{5}}(11)_s - \sqrt{\frac{3}{10}}(22)
 +\frac{1}{\sqrt{2}}(00)$ & $(11)_a$ \\
  & $\Xi N~(I=1)$ & $\sqrt{\frac{3}{5}}(11)_s+\sqrt{\frac{2}{5}}(22)$
  & $\frac{1}{\sqrt{3}}[-(11)_a+(30)+(03) ]$ \\
$-2$ & $\Sigma\Lambda$ & $-\sqrt{\frac{2}{5}}(11)_s
+\sqrt{\frac{3}{5}}(22)$ & $\frac{1}{\sqrt{2}}[(30)-(03)]$ \\
  & $\Sigma\Sigma~(I=0)$ & $\sqrt{\frac{3}{5}}(11)_s
-\frac{1}{2\sqrt{10}}(22)-\sqrt{\frac{3}{8}}(00)$ & $-$ \\
  & $\Sigma\Sigma~(I=1)$ & $-$ & $\frac{1}{\sqrt{6}}[2(11)_a+(30)+(03)]$ \\
  & $\Sigma\Sigma~(I=2)$ & $(22)$ & $-$ \\[1mm]
\hline
  & $\Xi \Lambda$ &  ${1 \over \sqrt{10}} [ (11)_s + 3 (22) ]$ &
${1 \over \sqrt{2}} [ -(11)_a + (30) ]$ \\
$-3$ & $\Xi \Sigma~(I=1/2)$ & ${1 \over \sqrt{10}} [ 3 (11)_s - (22) ]$ &
${1 \over \sqrt{2}} [ (11)_a + (30) ]$ \\
   & $\Xi \Sigma~(I=3/2)$ & $(22)$ & $(03)$ \\[1mm]
\hline
$-4$ & $\Xi \Xi~(I=0)$ & $-$ & $(30)$  \\
   & $\Xi \Xi~(I=1)$ & $(22)$ & $-$ \\
\end{tabular}
\end{center}
\end{table}

\end{document}